# Evidence for Dynamically Important Magnetic Fields in Molecular Clouds

Hua-bai Li[1*], Raymond Blundell[2], Abigail Hedden[2#], Jonathan Kawamura[3], Scott Paine[2] and Edward Tong[2]

[1]*Max-Planck Institute for Astronomy, Königstuhl 17, D-69117 Heidelberg, Germany*

[2]*Harvard-Smithsonian Center for Astrophysics, 60 Garden Street, MS-78, Cambridge, MA 02138, USA*

[3]*Jet Propulsion Laboratory, Pasadena, CA 91109, USA*

**Abstract**
Recent observational evidence that magnetic fields are dynamically important in molecular clouds, *compared to self-gravity and turbulence*, is reviewed and illustrated with data from the NGC 2024 region. One piece of evidence, turbulence anisotropy, was found in the diffuse envelope of a cloud (Av ≈ 1; Heyer et al. 2008); our data further suggests turbulence anisotropy in the cloud (Av >7) and even near the cloud core (Av~100). The data also shows that magnetic fields can channel gravitational contraction even for a region with super-critical $N(H_2)/2B_{los}$ ratio (the ratio between the observed column density and two times the line-of-sight observed field strength), a parameter which has been widely used by observers to estimate core mass-to-flux ratios. Although the mass-to-flux ratio is constant under the flux-freezing condition, we show that $N(H_2)/2B_{los}$ grows with time if gravitational contraction is anisotropic due to magnetic fields.

## 1. Introduction

How important a role magnetic fields play in molecular cloud and star formation has been long debated, largely because of observational difficulties. As summarized by McKee and Ostriker (2007), at that time observational data was thought to indicate that cloud cores were mostly close to the border of magnetically sub- and supercritical (however, see the discussion below in § 6.2), and that cloud-core turbulence and magnetic field energies are close to equipartition (Crutcher 1999a; Bourke et al. 2001). For the remaining, and majority part, of a cloud, our knowledge of magnetic fields was very limited. This was a significant problem, because cloud field strength is a crucial initial condition in some star formation theories (e.g. Mouschovias 1976, Shu et al. 1987) and in simulations (e.g. Price & Bate 2008; Heitsch et al. 2009), with significant influence on star-forming efficiency and rate.

Since 2007, our understanding of the strength of magnetic fields compared to turbulence and self-gravity has been remarkably improved because of new observations. Using NGC 2024 as an example, we summarize these improvements and show new evidence on: ordered cloud fields (§ 2); the correlation between field orientations and core shapes (§ 3); turbulence anisotropy (§ 4); and decoupling between turbulence and magnetic fields (§ 5). A discussion and summary are given respectively in the last two sections.

The importance of magnetic fields is best illustrated by a region which is highly turbulent and/or with strong gravitational energy. Here we choose NGC2024 for the following reasons:

(1) NGC 2024 is located in the complex of the Orion molecular region (OMC), which is a well-studied example of a turbulent, massive-star forming region. Compared to the relatively quiescent low-mass-star forming regions (e.g. Taurus molecular cloud (Goldsmith et al.

---

* li@mpia.de

# Current address: U.S. Army Research Laboratory 2800 Powder Mill Road Adelphi, MD 20783

2008) and Pipe nebula (Alves et al. 2008; Franco et al. 2010)), which have uniform magnetic fields revealed by optical polarimetry, the importance of magnetic fields in regions like OMC is much more controversial.

(2) Crutcher (1999) estimated the mass-to-flux ratios of 15 cloud cores. Among these, the NGC 2024 core has the highest value: 4.6 times of the critical value.

## 2. Galactic fields anchor in molecular clouds

It is hard to tell whether a cloud is globally super- or sub-Alfvenic by directly measuring the field strength. This is because the conventional methods, i.e. Zeeman measurements (e.g., Troland & Crutcher 2008) and the Chandrasekhar–Fermi method (e.g., Crutcher et al. 2004), cannot be used too far away from cloud cores. Secondly, both of these methods measure only the strength of some field components, either projections onto the line of sight or onto the plane of sky, and have significant uncertainty in their estimates (e.g. Hildebrand et al. 2009; Mouschovias & Tassis 2009, 2010; Crutcher et al. 2010; Houde et al. 2009). New methods have been proposed recently (Heyer et al. 2008; Li & Houde 2008), but they have not yet been widely applied.

We have tackled this problem in a different way. With flux freezing, it is not hard to imagine that a cloud core should not have any "memory" of the field direction outside the cloud, if the turbulence is so super-Alfvenic that the cloud fields are scrambled. This is seen in many numerical simulations, e.g. Ostriker et al. (2001) and Falceta-Goncalves et al. (2008). The field orientations in ICM (inter cloud media) and in cloud cores can be measured using, respectively, optical and submillimeter polarimetry. The uncertainty of field orientations is much smaller than that in field strength measurements, and projection effects will not enhance the apparent alignment between core and ICM fields, if there is any.

Fig. 1 shows the optical polarimetry detections within 50 pc (both on the sky and along the line of sight) of NGC 2024 (at distance 420 pc). The data is from the Heiles archive (2000). The mean direction is approximately 60 degrees NE. Using the RLT (Receiver Lab Telescope[#]), we have zoomed in on the 20' x 20' region centered at the peak (Fig. 2) with $^{12}$CO (7-6) emission. The RLT data was reduced with CLASS in the GILDAS software package (http://iram.fr/IRAMFR/GILDAS/"), and it shows two distinct regions. One is more diffused with Gaussian-like velocity profiles. The other is denser with clear self-absorption in the low-velocity wings (see more discussion in section 4).

Deep into the core-forming region, the core fields (Fig. 3) are mapped by the Hertz polarimeter (Dotson et al. 2010) at the Caltech Submillimeter Observatory. The mean core field direction is roughly at 50 degrees, which is very close to the field orientation detected by the optical data (Fig. 1) from a region 3 orders of magnitude larger and 5 orders of magnitude less dense.

Li et al. (2009) compared the fields from 25 cloud cores (sub-pc scale), from the Hertz (Dotson et al. 2010) and SCUpol (Matthews et al. 2009) archives, to their surrounding ICM (hundreds of pc scale) and found a significant correlation. Comparing with the cloud simulations in the literature, only sub-Alfvenic ones present a similar picture. Note that in these numerical simulations, the orderliness of the magnetic field is very sensitive to the Alfvenic Mach number ($M_A$). $M_A \approx 2$ is enough to make the field orientation random (see e.g. Falceta-Goncalves 2008, Ostriker et al. 2001, and Fig. 3 in Li et al. 2009).

Given the high correlation between cloud core and ICM fields, the field orientations in between,

---

[#] The RLT is a 0.8-meter THz telescope operated by the Submillimeter Receiver Lab of the Harvard-Smithsonian Center for Astrophysics in collaboration with Jet Propulsion Laboratory and the Universidad de Chile. Elevated at 5525 m on Cerro Sairécabur of the Atacama dessert, 40 km north of the ALMA site, the RLT is the highest ground-based radio telescope. This observation used the 810 GHz receiver system incorporating a HIFI Band-3 SIS mixer developed for Herschel Space Observatory (Hedden et al. 2009).

i.e. in the bulk cloud volumes, should be ordered. This was actually directly observed earlier: the field orientations from four GMCs were mapped using 450μm polarimetry (Li et al. 2006), and, except for the regions compressed by the $H_{II}$ bubbles, the fields are very ordered. However, the sample size of direct observations is hard to increase, because mapping the whole clouds is extremely time consuming.

## 3. Anisotropic gravitational contraction

Given that cloud fields and ICM fields are aligned, flux freezing should make a gravitational contraction anisotropic, and (somewhat) flattened high-density regions perpendicular to the mean field direction should be expected. Filamentary clouds perpendicular to the local ICM fields are widely observed, e.g. Heyer et al. (1987), McGregor et al. (1994), Rizzo et al. (1998), Pereyra & Magalhães (2004), and Alves et al. (2008). More examples are given in Fig. 1, where the elongated, highly-turbulent massive star forming clouds, including NGC 2024, are well aligned in a direction perpendicular to the mean field orientation.

Because cloud fields are ordered, the field/core-shape correlation is also expected, as observed in the NGC 2024 core region (Fig. 3). Fig. 3 covers the same region as Crutcher (1999) analyzed for NGC2024. He estimated that the mass-to-flux ratio ($N(H_2)/2B_{los}$) value here is highly supercritical (4.6 times of the critical value), the highest among the 15 regions for which he acquired measurements. Even so, the magnetic field still clearly channeled the contraction to form an elongated core with the long axis perpendicular to the field. See section 6.2 for discussion.

The observation of this kind of correlation, however, is not always straightforward, because the real core flatness is not always observable due to the projection effect. Observations suggest that core projections are not generally spherically shaped (e.g. Benson & Myers 1989) and that the most probable core shapes are nearly oblate (e.g. Jones et al. 2001). If the shortest axis of an oblate core indeed orients closely to the mean field direction, then the closer the line of sight is to being perpendicular to the shortest axis, the better the alignment which should be observed between the field projection and the short axis of the core projection. In other words, the more elongated the core projection, the better the alignment should be. This is exactly the trend Tassis et al. (2009) observed from 32 cloud cores, from the Hertz archive (Dotson et al. 2010).

## 4. Turbulence anisotropy

Given that cloud fields are ordered, turbulent velocities should also be anisotropic. Turbulent energy cascades more easily in the direction perpendicular to the mean field than in the direction aligned with the field, if the field is dynamically important compared to the turbulence. This is because the development of turbulent eddies will be suppressed in the direction parallel to the field. This phenomenon should be more prominent at smaller scales, where turbulent energy is lower and field curvature (and thus tension) will be larger if the field is bent.

An analytic model of anisotropic, incompressible MHD turbulence was first proposed by Goldreich & Sridhar (1995; 1997). Cho & Vishniac (2000), Maron & Goldreich (2001), and Cho, Lazarian, & Vishniac (2002) numerically verified the anisotropy predicted by the Goldreich-Sridhar model. Similar anisotropy was also observed in compressible MHD simulations from Cho & Lanzarian (2002) and Vestuto, Ostriker, & Stone (2003).

4.1 Cloud envelopes ($A_v \sim 1$ mag)
Heyer et al. (2008) first observed turbulence anisotropy in a molecular cloud by showing that the $^{12}CO$ (1-0) turbulence velocity spectrum derived by PCA (principal component analysis; Heyer & Schloerb 1997; Brunt & Heyer 2002) is steepest in the direction along the magnetic field. The region they studied is roughly 2° ×2° centered at 4h:50m, 27° (J2000) in the Taurus molecular cloud (TMC). The $A_v$ in this region is only around 1 mag (Schlegel et al. 1998), and thus the conclusion from this work is restricted to the cloud envelope.

Another related phenomenon Heyer et al. (2008) observed is that the $^{12}CO$ emission exhibits "streaks" that are aligned *along* the magnetic field direction. It is important not to mix the streaks here with the elongated density profile discussed in § 3. For higher density regions in the TMC, the elongated density profiles are still perpendicular to the field direction; for example, see the $^{13}CO$ map of Narayanan et al. (2008) and the $A_v$ = 4 mag contour shown by Kainulainen et al. (2009). The low-density streaks perpendicular to the high-density oblate regions are also "observed" in simulations with strong fields from, e.g., Price & Bate (2008) and Nakamura & Li (2008): while gas slides along the field lines to form a flattened high-density region *perpendicular* to the field because of anisotropic gravitational contraction, turbulence is also channeled by the field and, in the lower-density regions, results in streaks *aligned* with the magnetic field. These kinds of streaks are not seen in the simulations with weaker or no magnetic fields (Price & Bate 2008).

4.2 $A_v$ > 7 mag

Since the ICM field orientations are kept through clouds all the way down to cloud cores (§ 2), turbulence anisotropy should also be expected at higher densities. The RLT $^{12}CO$ (7-6) map (Fig. 2) of NGC 2024 is from a region within the $A_v$ = 7.2 mag contour in Fig. 1, so it has a much higher density than almost every position in TMC.

First we focus on the region that is less dense and with Gaussian-like line profiles (the dash-lined region in Fig. 2). A clump offset at (-420", 400") is elongated at roughly 40° NE (Fig. 2), which is only ~20° away from the ICM field direction. The coarse velocity resolution (~ 1 km/s) of the RLT backend prevents us from using the PCA method to study the turbulence spectra. Instead, we use a method modified from Li & Houde (2008) to study the correlation between the velocity dispersion (VD) and linear scales along different directions.

The method used by Li & Houde (2008) is based on how Ostriker et al. (2001) "observed" their simulations, and the validity has been recently double-checked by independent simulations (Falceta-Gonçalves et al. 2010; Tilley & Balsara 2010). In short, the lower envelope of a plot of VD versus linear scale can be used as an estimate of the turbulence velocity spectrum. Li & Houde (2008) obtained VDs from different scales by binning nearby pointings in the data cube to form beams of 1×1, 2×2, 3×3 … pixels. Where the pixel size is defined by the telescope spatial resolution and by pointing space. The idea is that the lower envelope keeps the line-of-sight scale to minimum, so that the VD variation from the lower envelope is mostly contributed from the varying beam size. We modify this method by binning nearby pointing to form beams of n×1 pixels, and the long axes are aligned with, for simplicity, ± 45° from the North (Fig. 2), i.e., close to perpendicular (⊥B) or parallel (//B) with the field direction at 60°. For each beam orientation and scale, we move the beam all over the region to get as many different pointings as possible. Then the median of the lower 25% of the VDs from each beam orientation at each scale is used as the estimate of the lower envelope at this particular scale. The result is shown in Fig. 4, a plot of $VD_{⊥B}$ versus $VD_{//B}$. The $VD_{//B}$ is symmetrically smaller. Moreover, the smaller the scale, the larger the VD difference, as one should expect for turbulence anisotropy caused by magnetic fields.

The velocity profiles from the higher density region in Fig. 2 outside the dash-lined region are dominated by expansion, instead of turbulence. The closer to the emission peak, the more of the low-velocity wing is self-absorbed. When these spectra are Gaussian fitted, the Gaussian peak velocity increases with the peak intensity, as illustrated by the Gaussian peak-velocity gradients in Fig. 2. This is a signature of cloud expansion (Evans 1999; Lee & Kim 2009). In their simulations, Lee & Kim (2009) showed that line profiles from a core-forming turbulent cloud can be complicated: while the core is contracting, the outer regions show not only contracting but also sometimes expanding motions (see also Keto et al 2006; Gao & Lou 2010). Observations also show dichotomy in the collapse/expansion kinematics state from massive-star forming cores (Velusamy et al. 2008).

4.3 Near the core ($A_v \sim 100$ mag)

As shown by the vectors in Fig. 3, the field direction in the NGC 2024 core region is inherited from the local ICM. On September 4th and 8th 2008, the SMA (Submillimeter Array) was pointed about 1' away from the FIR3 toward the NE and SW, centered at respectively ($05^h41^m46^s.0$, $-01°53'12".0$) and ($05^h41^m41^s.5$, $-01°54'57".0$) (J2000), to check for streaky footprints of turbulence anisotropy. The tracer used is $HCO^+$ (3-2), whose critical excitation density is approximately $n(H_2) = 10^5/cm^3$, which is equivalent to $N(H_2) = 10^{23}/cm^2$ in this region (Crutcher 1999) and $A_v \sim 100$ mag (Harjunpaa 2004).

The result is shown by the contours in Fig. 3. Elongated structures aligned with the field and perpendicular to the flattened core region are clearly seen. The streak in the SW bends in the same way as the field lines. We can tell from the field-line density (Li et al. 2010) that the field is stronger in the SW than NE, and this is also consistent with the field strength estimated by the Zeeman measurements from Crutcher et al. (1999). The fact that the SW streak is more elongated agrees with the relative field strengths.

The slimness (comparing with the telescope resolution) of the streaks, however, prevents us from using PCA or the method in § 4.2 to study the velocity anisotropy.

## 5. Decoupling between magnetic fields and turbulence

So far we have assumed that flux freezing is a good approximation. Moving toward small scales, however, the decoupling of neutral flows from magnetic fields and plasma becomes inevitable. This process is important because the friction between decoupled ions and neutrals may accelerate the dissipation of turbulent energy (e.g. Zweibel & Josafatsson 1983). Direct observation of the different velocities from decoupled ions and neutrals is difficult, because the decoupling may happen at a scale well below the spatial resolutions of current telescopes. However, Li & Houde (2008) showed that this decoupling causes differences between the VDs of coexistent ions and neutrals, even with telescope beam sizes larger than the decoupling scale. This is because a VD results not only from contributions of turbulent eddies with scales comparable to the beam size, but also from all the smaller eddies contained within it. Since field lines and ions are always coupled (through Lorentz force and electric force between ions and electrons trapped by field lines) and because Alfven waves are damped at a scale larger than the scale at which hydrodynamic viscosity sets in (Li & Houde 2008), ions, compared to neutrals, should have a larger cutoff scale of the turbulent energy spectrum. So ions are predicted to have smaller VDs than coexistent neutrals.

Besides the $HCO^+$ data in Fig. 3, we also observed HCN (3-2) from the same regions. $HCO^+$ and HCN is the ion/neutral pair with the highest spatial correlation (e.g. Lo et al. 2009), and is thus ideal for the VD comparision. The detailed data analysis is presented in a separate article (Li et al. 2010); here we summarize the result: the systematic difference between the VDs of ions and neutrals is larger in the SW, where the field is stronger (§ 4.3). Li et al. (2010) discuss various possible causes for this VD difference. However, given that the narrower $HCO^+$ lines are optically thicker, slightly more spatially extended, have more unresolved hyperfine structures than HCN, and are away from the outflows in NGC 2024, decoupling between field lines and turbulent is the only plausible mechanism for the VD difference.

Besides NGC 2024, Li et al. (2010) also demostrate the correlations of field-strength and VD-difference in molecular clouds M17 and DR21(OH). Houde et al. (2000 a, b) have shown that this ion/neutral VD difference is a general condition in the 10 clouds they observed.

**6. Discussion**

6.1 The "hourglass"

The hourglass-shaped field morphologies (e.g. Girart et al. 2006; Girart et al. 2009) from cloud cores have been popular as evidence of magnetic fields energetically dominating turbulence. The field shape indicates that they are ordered prior to the "pinch" and that magnetic field is the main force against the core contraction under self-gravity.

The same conclusion cannot be derived for scales larger than an hourglass. Because, for example, assuming a cloud is super-Alfvenic, fields shaped by turbulent shocks are also ordered. Turbulence loses energy to magnetic fields while compressing them, until the magnetic pressure and/or tension grows strong enough to resist the compression. Though the smaller, less energetic eddies within the shocks cannot tangle the post-shock fields and thus a collapse will lead to an hourglass, the turbulence is still super-Alfvenic at cloud scale in this case. So hourglass-shaped field morphologies from cloud cores cannot distinguish between super- and sub-Alfvenic at larger scales. Cloud-scale fields are very difficult to observe (§ 2), and the existence of large-scale hourglasses (Mouschovias 1976) cannot yet be well tested.

However, post-shock field orientations should not correlate with mean field direction from larger scales in a super-Alfvenic cloud. This is why the multi-scale comparison in field orientations, as described in § 2, is important.

6.2 Dynamically important fields in magnetically "supercritical" clouds

As described in § 1, cloud cores are thought mostly to be on the border between magnetically sub- and supercritical, mainly based on Crutcher (1999a) and Bourke et al. (2001). Thus, some claims are made in the literature that magnetic fields cannot effectively regulate gravitational collapse, because of the lack of sub-critical cores (e.g. Mac Low & Klessen 2004). Before one can reach this conclusion, however, some other facts should be considered:

(1) Given a field strength and a mass density, there always exits a scale above which the system is supercritical. This is because magnetic flux depends quadratically on scale, while mass is cubic. If cloud cores stem from this "magnetic threshold", of course they should be just supercritical.

(2) There are both observations (e.g. Crutcher 1999**)** and simulations (e.g. Burkhart 2009) suggesting that cores are close to equipartition between magnetic and turbulence energies. We showed that (§ 2; Li et al. 2009) clouds have to be sub-Alfvenic to explain the core-ICM field correlation. So magnetic fields are at least as important as turbulence in both cloud and core scales.

(3) Crutcher (1999) used $N(H_2)/2B_{los}$ to estimate the mass to magnetic flux ratio, $M/\Phi_B$. Assuming flux freezing, $M/\Phi_B$ is time independent. However, here we show that $N(H_2)/2B_{los}$ can increase while clouds contract under self-gravity.

Consider a uniform sub-critical cylindrical volume with diameter $R$, height $L$, and magnetic field B in the direction aligned with the cylinder axis (Fig. 5). $L$ will decrease under self-gravity collapse, while R stays almost constant comparing to $L$ because the cloud is sub-critical. $M/\Phi_B = N(H_2)_0/B$ also stays constant, where $N(H_2)_0$ is the column density when the line of sight is parallel with B. Given a line of sight $\theta$ degrees from the field direction (Fig. 5):

$$\frac{N(H_2)}{2B_{los}} = \frac{N(H_2)_0 A/A_p}{2B\cos\theta} \qquad [1]$$

Where A is the area of the cylinder circular cross-section, and $A_p$ is the projection area of the cylinder along the line of sight:

$$A_p = RL\sin\theta + \pi(\frac{R}{2})^2 \cos\theta \qquad [2]$$

$N(H_2)/2B_{los}$ versus $L$ from some sample sight lines are shown in Fig. 5, where $N(H_2)/2B_{los}$ is normalized to the value when $L=R$. Fig. 5 shows that the ratio $N(H_2)/2B_{los}$ of a sub-critical volume increases with time, which may make it appear super-critical, especially in the later stages of a core formation, when the core shape becomes oblate due to the anisotropic collapse (section 3). This is not hard to visualize: as long as the line of sight is not parallel to the cylinder axis, $N(H_2)$ increases with the decreasing $L$, but $B_{los}$ is independent of $L$. So $N(H_2)/2B_{los}$ tends to overestimate $M/\Phi_B$ for a sub-critical volume.

This effect also works, though not as prominently, on super-critical volumes, for which $R$ decreases much faster. As long as $R$ does not decrease as fast as $L$ because of the magnetic pressure (anisotropic contraction; Mouschovias 1991), $A/A_p$, and thus $N(H_2)/2B_{los}$, increases with time.

We note that we are talking about an effect different from the projection effect on $N(H_2)$ discussed by Bourke et al. (2001). What Bourke et al. did is equivalent to discussing the effect of θ assuming L=0. We are emphasizing that $N(H_2)/2B_{los}$ can grow significantly as L decreases.

## 7. Summary

With NGC 2024 as an example, we illustrated recently-found evidence for dynamically important magnetic fields in molecular clouds, including ordered cloud fields (section 2); anisotropic gravitational contraction (field/core-shape correlation; section 3); anisotropic turbulence in cloud envelopes (4.1); and turbulence/field-lines decoupling (section 5). The NGC 2024 data further provides new evidence, indicating that

(1) Highly anisotropic gravitational collapses can happen even in a region with highly magnetically supercritical $N(H_2)/2B_{los}$ value (section 3); and
(2) Turbulence anisotropy happens with $A_v > 7$ mag (4.2), a typical cloud density, and even to $A_v \sim 100$ mag near the cloud core (4.3).

Since strong fields may boost the observed $N(H_2)/2B_{los}$ values by channeling gravitational contraction (6.2), claiming that magnetic fields are not dynamically important just because of the high $N(H_2)/2B_{los}$ values (especially in a later stage of core formation) ignores the significant geometrical effects of flattening (section 3), which is itself a consequence of the presence of dynamically important magnetic fields. Similarly, strong turbulence can boost local field orderedness (6.1), so multi-scale observations (section 2) are needed for using field morphology to distinguish between globally super- and sub-Alfvenic molecular clouds.

Our deepest appreciation goes to the staff of the SAO Submillimeter Receiver Lab: Robert Kimberk, Steven Leiker, Cosmo Papa, Patricia Riddle, and Michael Smith. We are grateful for the support of RLT observations from Jorge May, Leo Bronfman, Claudio Barrientos, Marcos Díaz, José Donoso, Daniel Lühr, Walter Max-Moerbeck, Daniel Marrone, and Kevin Rauch. We thank Zhi-Yun Li, Matthew Kunz and the referee for many insightful comments. H.L.'s research is supported by the postdoctoral fellowships from Max-Planck-Institut für Astronomie and from Harvard-Smithsonian Center for Astrophysics. The RLT was supported in part by Smithsonian Institution internal research and development funds. Part of the RLT research was carried out at the Jet Propulsion Laboratory, California Institute of Technology, under a contract with the

National Aeronautics and Space Administration. Operation of the RLT on Sairecabur has been carried out in collaboration with the Universidad de Chile. The Submillimeter Array is a joint project between the Smithsonian Astrophysical Observatory and the Academia Sinica Institute of Astronomy and Astrophysics and is funded by the Smithsonian Institution and the Academia Sinica.

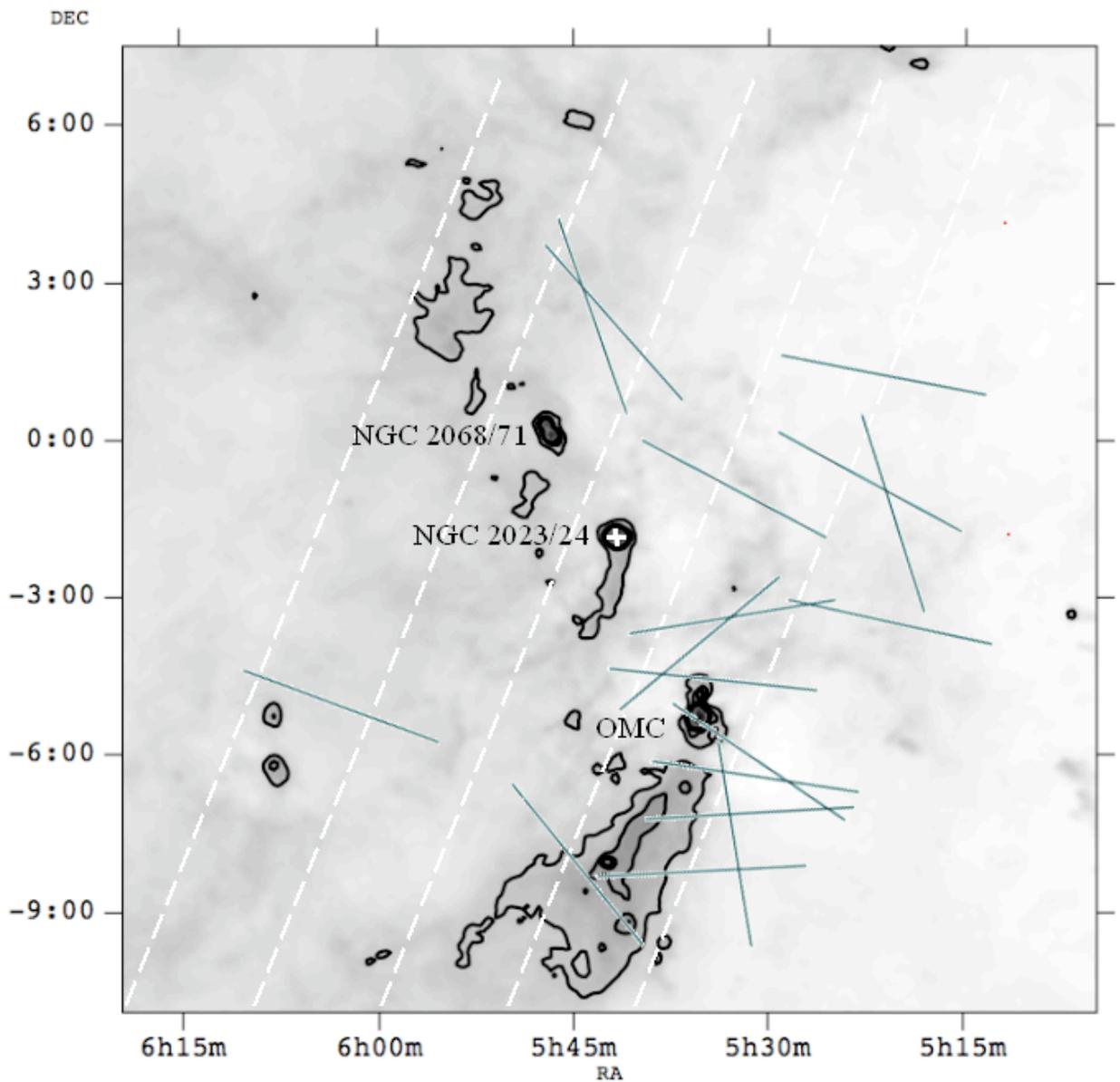

**Fig. 1** A dust column density map overlapped by optical polarimetry data within 50 pc from NGC 2024. The contour levels are 1.2, 3.2, 5.2 and 7.2 mag (Schlegel et al. 1998). The mean field direction based on the polarimetry data is around 60° NE. The dashed lines are in a direction perpendicular to the mean field direction. Note how well the elongated cloud shapes are aligned with the dashed lines. Fig. 2 is a zoom-in of the 20' x 20' region centered at the white "+".

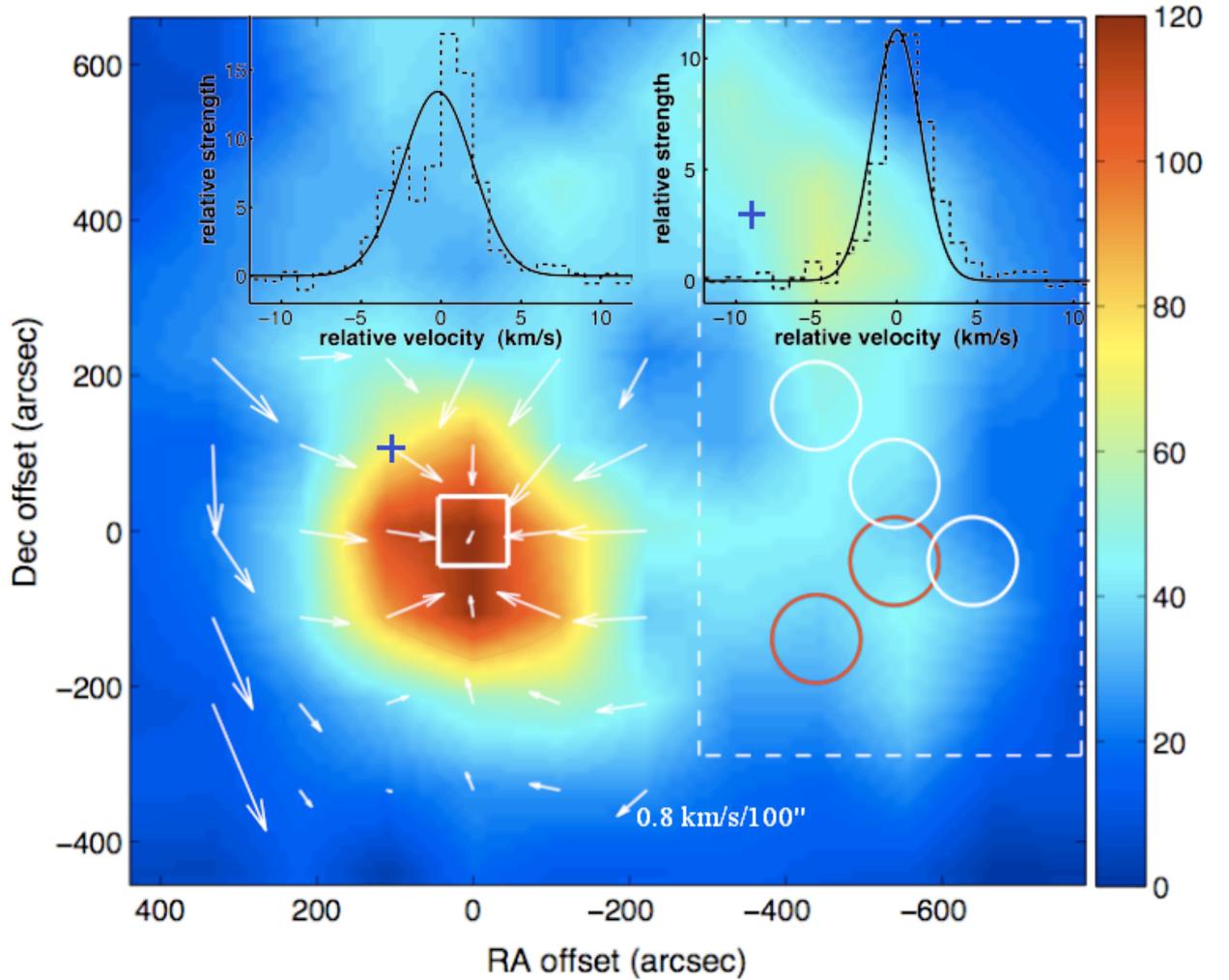

**Fig. 2** A $^{12}$CO (7-6) map of NGC 2024; the relative strength of the zero-moment is presented by the false colors. The offset is relative to 5h41m43s, -1°54'22" (J2000). The left (right) line sample (dark dashed line) is from the position noted by the left (right) blue "+". Spectral lines from the lower left region are self-absorbed in the low-velocity wings. The higher the peak intensity of a spectrum, the stronger the self-absorption at low velocity, so the peak velocity from a single-Gaussian fit (dark lines) increases with the intensity. This is illustrated by the peak velocity gradients of the Gaussian fits (the arrows; note that the directions represent the variation of the self-absorption profile, not velocity directions); the arrow lengths are proportional to the velocity gradients. This is a signature of expansion (see § 2). Spectral lines from the region within the white dashed lines are Gaussian-like. The circles represent the RLT resolution (115") at 810 GHz and pointing spacing (100"). The red (white) circles together stand for a binned telescope beam elongated at 45° NW (NE), a direction which is close to be perpendicular (parallel) to the local ICM fields (Fig. 1). These binned beams are used to study the scale/velocity-dispersion correlations along ± 45° from the North (see § 4). The results are shown in Fig. 4

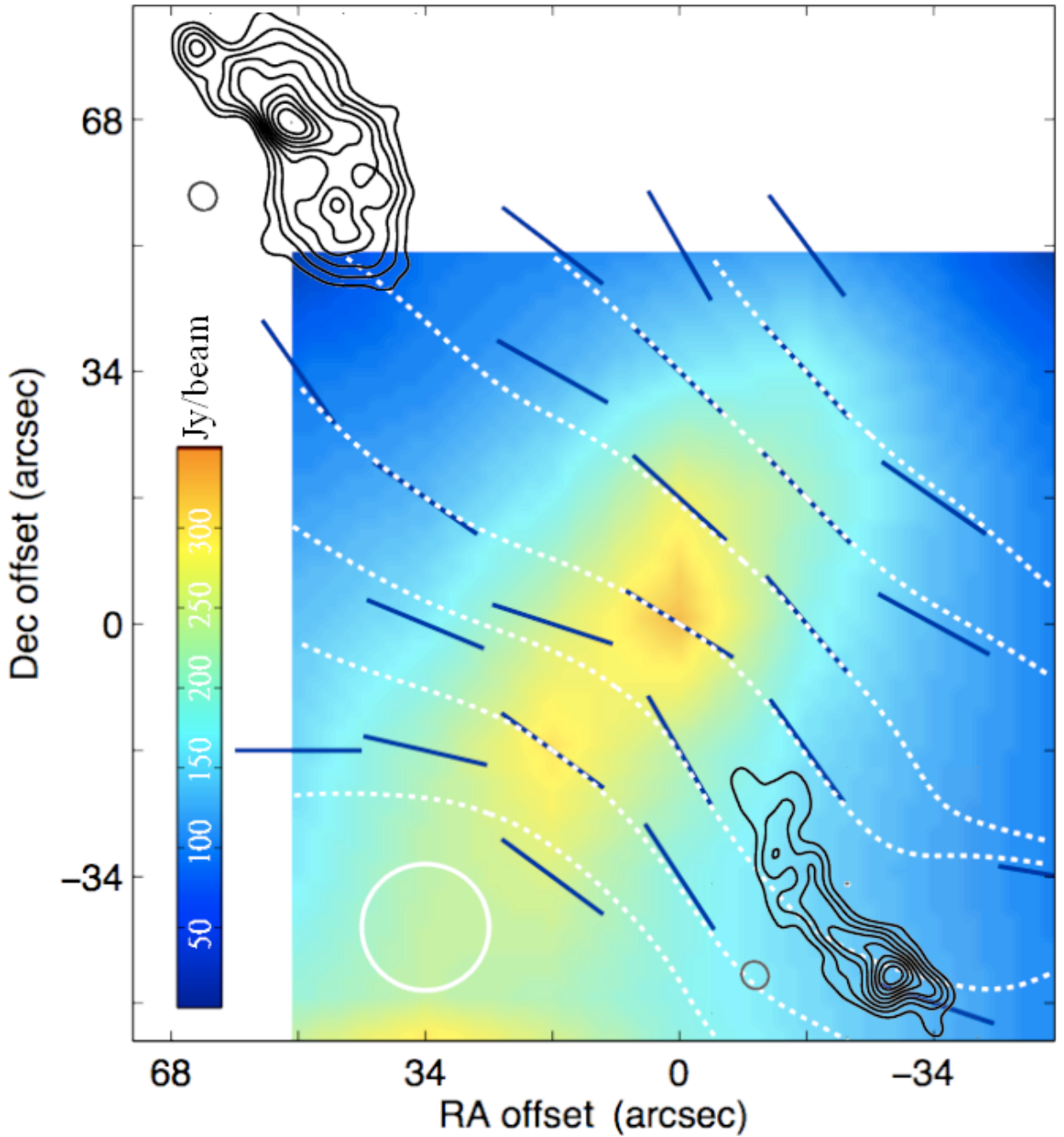

**Fig. 3** False-color map: dust thermal emission at 350μm from the NGC 2024 cloud cores (Dotson et al. 2010) with 20″ resolution (white circle), centered at FIR 3 (5h41m43s, -1°54'22" J2000; the origin). Blue vectors: the magnetic field directions inferred from the 350 μm polarimetry data (Dotson et al. 2010). Note that the mean field direction here is only ~ 10° from the mean field direction in Fig. 1, and that how different are the densities and scales between Fig. 1 and Fig. 3. Dotted lines: magnetic field lines based on the 350 μm polarimetry data (Li et al. 2010). Contours: HCO+ (3-2) zero-moment maps (Li et al. 2010) with ~ 3.7″× 3.4″beams (the ovals SE from the contours). The highest contour level has 90% of the peak intensity, and the following levels decrease linearly by 10 % of the peak value. Note that the dense dust ridge is perpendicular to the mean field direction and that the HCO+ streaks are parallel with the fields (see § 3 and § 4).

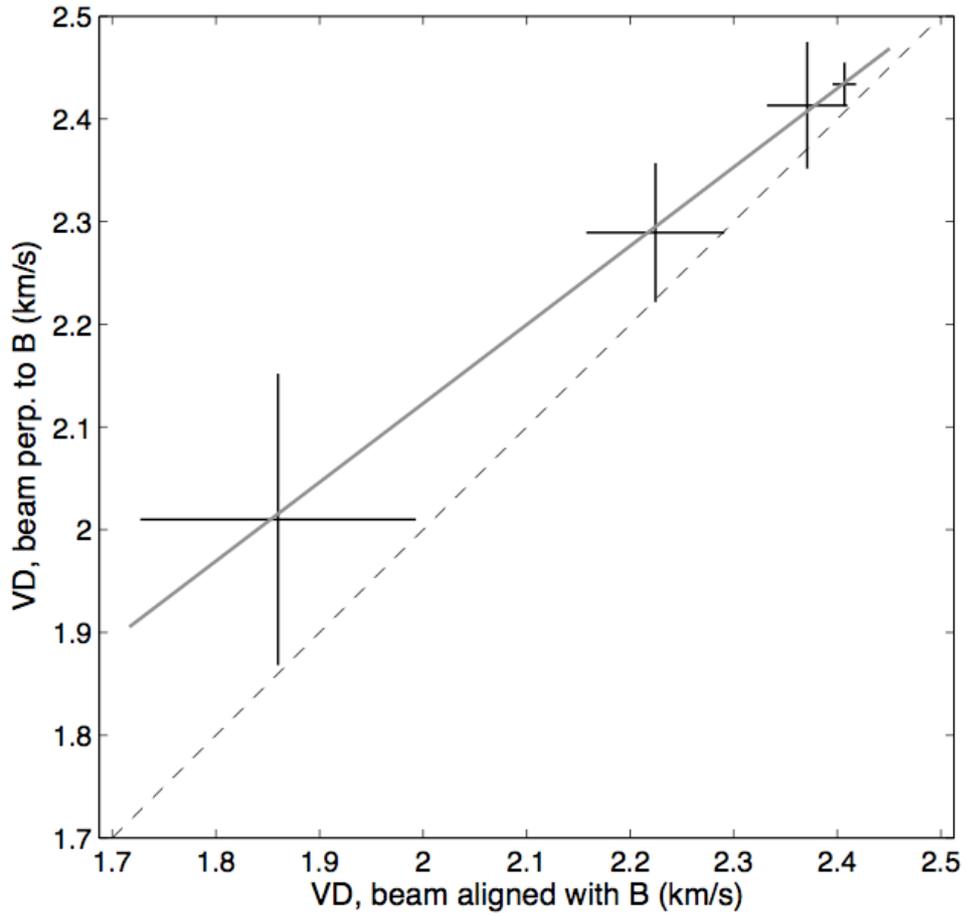

**Fig. 4** The lower envelope of a plot of velocity dispersions (VD) versus scales (L) works as an estimate of the turbulence velocity spectrum (§ 5; Li & Houde 2008). Here we compare the lower envelopes from two VD-L plots at various scales: one with L varying along the direction at 45° NE (x-axis) and the other at 45° NW (y-axis). From left to right, the data points are from respectively L = 2, 3, 4, and 5 times of the telescope resolution (see § 4). We estimate the lower envelope at each scale using the median of the lower 25% of the VDs at this scale. The error bars denote the interquartile range for the lower 25% VDs. The solid line fitted to the data, and the dashed line for "y = x" illustrate the fact that the VD difference is more prominent at smaller scales.

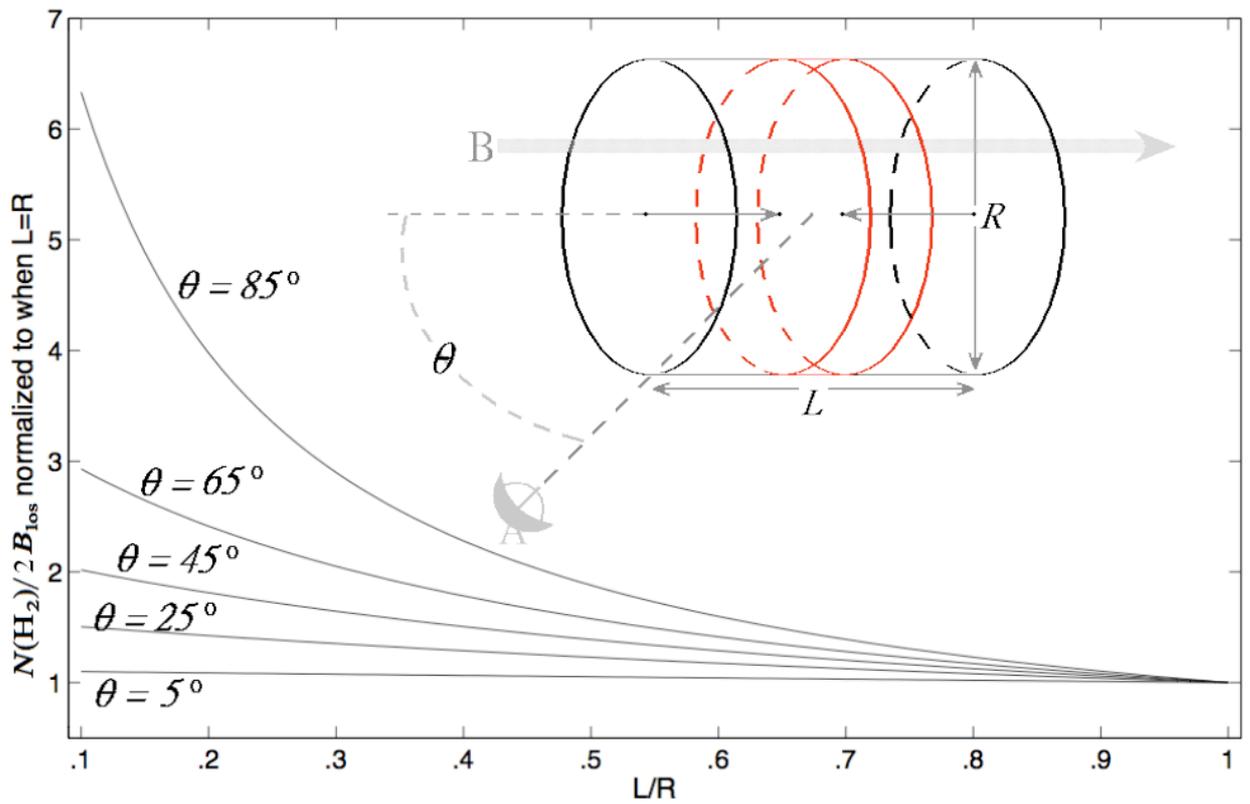

**Fig. 5** This figure illustrates that the ratio of column density to two times of the strength of line-of-sight field ($N(H_2)/2B_{los}$), which has been used to estimate constant mass-to-flux ratio (e.g. Crutcher 1999), can increase with time. The magnetic field is parallel with the axis of the uniform cylindrical sub-critical volume, with length $L$ and diameter $R$. $L$ is contracting because of self-gravity, while $R$ is assumed constant because of the volume is sub-critical. Assuming that $L$ initially equals $R$, the plot shows how $N(H_2)/2B_{los}$ varies with $L$ observed along various lines of sight. The angle between a line of sight and the cylinder axis is $\theta$.